\documentstyle[epsf]{l-aa}     
\begin{document}
\title{HIPPARCOS calibration of the peak brightness of four SNe Ia and the value of $H_{0}$}
    \author{P. Lanoix\inst{1,2}}
    \offprints{P. Lanoix}
    \thesaurus{08.19.4; 12.04.3  } 
    \institute{
              CRAL - Observatoire de Lyon,\\
              F69230 Saint-Genis Laval, FRANCE,\\
 \and
              Universit\'e Claude-Bernard-Lyon1\\
             F69622 Villeurbanne, FRANCE\\
           }
  \date{Accepted: November 06, 1997}
 
 \maketitle
 
   \begin{abstract}
HIPPARCOS geometrical parallaxes allowed us to calibrate the Cepheid Period-Luminosity
relation and to compute the true distance moduli of 17 galaxies. \linebreak Among these 17
 galaxies, we selected those which \linebreak generated type Ia Supernovae (SNe Ia).
We found NGC 5253, parent galaxy of 1895B and 1972E, IC 4182 and NGC 4536
parents of 1937C and 1981B, respectively. \\
We used the available B-band photometry to determine the peak brightness of
these four SNe Ia. We obtained $\langle M_{B}(MAX)\rangle = -19.65 \pm 0.09$.
 Then, we built a 
sample of 57 SNe Ia in order to plot the Hubble diagram and \linebreak determine its
zero-point. Our result ($ZP_{B} = -3.16 \pm 0.10$) is in agreement with 
other determinations and allows us to derive the 
following Hubble constant :\linebreak $H_{0} = 50 \pm 3~(internal)~km.s^{-1}.Mpc^{-1}$.

   \end{abstract}

\section{Introduction}
A Supernova represents a sudden brightening of a star by about 20 magnitudes. 
These events have very bright absolute maximum magnitudes which allow us to
detect them up to cosmological distances. Moreover, in the light of 
our knowledge of spectral type Ia Supernovae, their peak brightness are 
supposed to have a small dispersion, thus they are less sensitive to
Malmquist bias, and SNe are supposed to be not
 significantly affected by peculiar motions (because they are situated remote
enough to make the corrections from the 
Virgo infall less uncertain).
One early proof was produced when Kowal (1968) plotted the Hubble diagram 
($m_{MAX}$ vs redshift) for some SNe Ia and highlighted its small dispersion
and its linearity. Many studies claimed that the intrinsic dispersion may be
 less than 0.3 mag (assuming the exclusion of some abnormals events).\\
Type Ia Supernovae are thus considered to be excellent cosmological distance indicators (``standard candles'') and
 provide us a very useful tool to estimate the Hubble constant, as long
we can independently calculate their absolute magnitude.  \\

Our present goal is to calibrate the maximum peak brightness of SNe Ia thanks
 to a previous work where we have determined
the distance moduli of 17 galaxies based on the geometrical calibration
of the Cepheid Period-Luminosity relation.
We checked those which have generated a
SN Ia and found only three galaxies lodging  four SNe Ia : 1895B and 1972E in 
NGC 5253, 1937C in IC 4182 and 1981B in NGC 4536.   
Then using both the three galactic distance moduli and the B-band photometry 
(as homogeneous as possible)
of the four SNe Ia, we are able to compute the mean absolute magnitudes at maximum.\\
We make a selection from among a large sample of distant SNe Ia to build a 
reliable sample and plot a
 Hubble diagram whose zero-point is computed. We finally use both the peak brightness and the zero-point values to derive the Hubble constant.

\section{The parent galaxies}
The data presented in this section are summarized in \linebreak Table \ref{galaxy-sn}.\\

\subsection{The distance moduli}
We use in the present paper the results of a previous work (Paturel et al.,
 1997a) where new distance moduli were obtained for 17 calibrating galaxies.
In order to derive this moduli, we have defined a new calibration of the 
Period-Luminosity relation (independently of previous determinations) from 
new geometrical parallaxes of galactic Cepheids obtained with the 
HIPPARCOS satellite. This new calibration was combined with a compilation
of extragalactic Cepheid measurements in the BVRI photometric system. \\
The external error of the new zero-point is about 0.25 mag in absolute 
magnitude. Considering that this error is much larger than
the others when computing the distance moduli ($\sigma_{m}, \sigma_{\log P}$, etc...), we will assume that the
external error on the moduli is about 0.30 mag. Even though the precision of this result is poor, its accuracy 
(the measure of how close the result is to the true value) 
is quite acceptable.  \\
In the following, we will quote the external errors along with the external
 errors in order to appreciate the effect of both error sources,
and we will 
compute both the internal and the external error on the Hubble constant.    

\subsection{NGC 5253}
NGC 5253 is the host galaxy of 1895B and 1972E. According to our previous results
based on the HIPPARCOS geometrical calibration (Paturel et al., 1997a)
 we assign to this galaxy a true distance modulus :
$$\mu   =  (m - M)_{0} 
      =     27.96 \pm 0.05~(internal)$$
According to Burstein \& Heiles (1984), the B-band galactic
extinction in its direction is $A_{g} = 0.20 $.
The morphology is uncertain but it is probably a spiral (morphological type code 7-8 according to the
LEDA database) and corresponds to a kind of S type. It's a galaxy 
whose total color is $(B-V)_{T}^{0}$ = 0.26 (de Vaucouleurs et al., 1991).

\subsection{IC 4182}
IC 4182 is the parent galaxy of 1937C. It is a Sm type galaxy (morphological type code
8-9), and the B-band galactic 
extinction in its direction ($A_{g}$) is negligible. Its true distance 
modulus is supposed to be : $$ \mu = 28.50 \pm 0.03~(internal)$$

\subsection{NGC 4536}
NGC 4536 is the parent of the more recent SN, 1981B. Its morphological type is
SBbc (morphological type code = 4), and the B-band galactic 
extinction in its direction ($A_{g}$) is negligible too. 
We assign to it a true distance modulus~:
$$ \mu = 31.18 \pm 0.03~(internal)$$ 
Its total color $(B-V)_{T}^{0}$ equals 0.47.

\begin{table*}
\begin{tabular}{|l|l|l|l|l|l|}
\hline
R. A. 2000 DEC.  &Name   & Type  &Ag      & SN Ia        & $\mu$\\
h mn s deg ' ''    &       &       &        &              &    \\
\hline
$133955.8-313841$&NGC5253& S?    &0.20    & 1895B, 1972E & 27.96$\pm$0.05\\
$130549.3+373621$&IC4182 & Sm    &0.00    & 1937C        & 28.50$\pm$0.03\\
$123426.9+021119$&NGC4536& SBbc  &0.00    & 1981B        & 31.18$\pm$0.03\\    
\hline
\hline
\end{tabular}
\caption{Host galaxy information.}
\label{galaxy-sn}
\end{table*}

\section{Supernovae Ia photometry}
\subsection{1895B}
The discovery was made by Miss Fleming on 1895 July 8, at Arequipa
 (Fleming \& Pickering, 1896). The star was located 23 arcsec North of NGC 5253.
The light curve was plotted later by Hubble \& Lundmark (1922) 
using subsequent observations made at Harvard observatory. Walker (1923) put the observations
on a revised $m_{pg}$ scale and derived $m_{pg}$(max) = 8.0. 
However Leibundgut et al. (1991) fitted a type Ia 
template to the light curve (even though only loose constraints can be placed
on it) and obtained $m_{pg}$(max) = 7.03 (removing their extinction 
$A_{pg}$ = 0.13). However some evidences can explain 
that this result is wrong (see Saha et al., 1995, for details) and that 
$m_{pg}(max) = 8.05 \pm 0.17 $ seems to be a good compromise solution.\\
Following the $m_{pg}$ tranformation to B system of Arp (1961), and assuming
$(B-V)_{B(max)} = +0.09$ (Sandage \& Tammann, 1993) we finally obtain :
$$B(max) = 8.33 \pm 0.20$$ \\
Schaefer \& Bradley (1995) scanned the old SN plates and derived 
$B(max) = 8.26 \pm 0.11$ using the most likely shape, which is 
consistent with the previous value (note that they placed conditions on
 a firm limit on the peak magnitude $B(max) < 8.49 \pm 0.03$).\\
However their result may appear to be unreliable due to the 
age of the observations and to
the many transformations needed. \\
We can now calculate the peak absolute magnitude which is simply :

\begin{eqnarray}
M_{B}(MAX)&  = & B(max) - \mu - A_{g} \nonumber\\
& =  &   -19.83 \pm 0.23~(internal)\\
& =  &   -19.83 \pm 0.37~(external) \nonumber
\label{M1895B}
\end{eqnarray}
where $A_{g}$ is the B extinction according to Burstein \& Heiles (1984).

\subsection{1972E}
The discovery was made by Kowal on 1972 May 13 (Kowal, 1972), the 
observations began on May 17 at the European Southern Observatory 
in Chile and were conducted by 
Ardeberg \& de Groot (1973). 1972E was located 56 arcsec West and 85 arcsec South
of the nucleus of NGC 5253. It appeared to be a prototype of SNe Ia and was actually
 used to define the type Ia. Leibundgut et al. (1991) fitted a type Ia
template to the well constrained light curve, and obtained 
(removing their extinction $A_{B}$ = 0.13) :
$$B(max) = 8.58 \pm 0.10$$  
So that we obtain :
\begin{eqnarray}
 M_{B}(MAX)& = & -19.58 \pm 0.15~(internal) \\
           & = & -19.58 \pm 0.33~(external) \nonumber 
\label{M1972E} 
\end{eqnarray}

\subsection{1937C}
The discovery was made by Baade \& Zwicky (1938). The photographic photometry 
was made on Palomar Mountain and
 was very accurate. 1937C is also a prototype a
SNe Ia like 1972E. The Leibundgut et al. (1991) fit seems very reliable and 
results in $m_{pg}(max) = 8.50 \pm 0.05 $ (removing their extinction $A_{B}$ = 0.13).
 Assuming the same transformation as
before (Arp, 1961) with $(B-V)_{B(max)} = +0.19 \pm 0.15$ (Saha et al., 1994),
a further correction of 0.07 mag is required because 
this photometry produces brighter results than the
photoelectric one (Saha et al., 1994), leading us to :
$$B(max) = 8.83 \pm 0.11$$ And : 
\begin{eqnarray}
 M_{B}(MAX) & = & -19.67 \pm 0.15~(internal) \\
            & = & -19.67 \pm 0.33~(external) \nonumber   
\label{M1937C} 
\end{eqnarray}

\subsection{1981B}
On 1981 March 2, Tsvetkov discovered a 12th magnitude SN Ia (Aksenov, 1981)
located 36 arcsec east and 36 arcsec north of the nucleus of NGC 4536.
The light curve is very well constrained and
the Leibundgut et al. fit (in agreement with Phillips, 1993, and Schaefer, 1995)
leads to 
$B(max) = 12.00 \pm 0.10 $
, and then :
$ M_{B}(MAX) = -19.18 \pm 0.14 $.\\

Although this SN seems obviously to be less luminous than the three 
others, all studies considered it as a completely normal event.
The recent revised results from Patat et al. (1997) suggest the value 
$B(max) = 11.74 $. Such a result would lead to $ M_{B}(MAX) = -19.44 $.\\
However we may also consider some evidence of a certain color excess
of about $E_{B-V} \simeq 0.10 \pm 0.05$ (Branch et al., 1983, 
Buta \& Turner, 1983, Saha et al., 1996) even though it is not well constrained.
  Burstein \& Heiles claimed $A_{g}$ is
negligible for NGC 4536, but 1981B may be extinguished inside its host galaxy contrary
to 1937C which has no color excess.
 We would then obtain $A_{B} \simeq 0.41$ (with a
ratio of total to selective absorption of 4.1 for the photometric
B-band, Savage \& Mathis, 1979), and  
finally from the dereddened magnitude :
\begin{eqnarray}
M_{B}(MAX) & = & -19.59 \pm 0.23~(internal) \\
           & = & -19.59 \pm 0.38~(external) \nonumber
\label{M1981B}
\end{eqnarray}
 in better agreement with previous determinations.\\
 
\subsection{The mean value}
From equations \ref{M1895B} to \ref{M1981B}, assuming :
\begin{eqnarray}
\sigma=\frac{1}{\sqrt{\sum{\frac{1}{\sigma_{i}^{2}}}}} \nonumber
\end{eqnarray}
the peak brightness weighted 
mean value\footnote{The weight is taken as the inverse of the 
square of individual standard error} of our four SNe Ia is :
\begin{eqnarray}
 \langle M_{B}(MAX)\rangle & = & -19.65 \pm 0.09~(internal)
\label{Mmean}
\end{eqnarray}
\begin{eqnarray}
~~~~~~~~~~~~~~~~~~~~ & = & -19.66 \pm 0.18~(external) 
\label{Mmean2}
\end{eqnarray}
We could compute this mean value using only part of the data, because 
one can argue that some are less reliable. However it would not change 
significantly  the result 
because the four values are quite slightly scattered and the computed mean
is very close to the more reliable values (we would obtain 
$\langle M_{B}(MAX)\rangle = -19.62 \pm 0.11$ using only 1937C and 1972E).

\section{The Hubble diagram and $H_{0}$}
\subsection{The data}
We needed a database of Supernovae to construct our Hubble diagram :
we used the update version (online version, http://athena.pd.astro.it/$\sim$supern/snean.txt)
of the Supernovae Asiago Catalogue (Barbon et al., 1989), 
which contains 1130 Supernovae at the moment, 
from 1885 to 1997. \\
 First of all, we selected the type Ia according to this
catalogue and we removed those without B-band photometry and 
with unidentified host galaxies.  
Then we used the Lyon-Meudon Extragalactic Database (LEDA, 
http://www-obs.univ-lyon1.fr/leda/leda-consult.html) in order to find the  
radial velocity of each host galaxy and the galactic extinction in the
B-band from Burstein \& Heiles. We selected the velocities 
corrected from the infall of the Local Group towards Virgo ($v_{Vir}$) according to 
Paturel et al. (1997b), where the chosen infall velocity of the Local Group is
$170~km.s^{-1}$ (Sandage \& Tammann, 1990).
This last step removed
some more SNe Ia whose parent galaxies had no velocity measurements or
no galactic extinction (two occurences).  
It appears that the SN 1963I type may be uncertain. Following 
Leibundgut \& Tammann (1990), we excluded it because there is no
evidence available of its classification as type Ia SNe.  \\

We then checked the 57 remaining SNe Ia in order to  
estimate the errors on the maximum magnitudes. We assign a typical uncertainty
of 0.14 mag to the best extinction corrected magnitudes, taking into
account both the measurement itself ($\sigma = 0.1 $) and the 
extinction correction ($\sigma = 0.1$). This
uncertainty concerns 37 SNe Ia (plus 1972E).\\

We found three  no ``Branch normal'' SNe Ia in our sample
 (Branch \& Miller, 1993, Branch et al., 1996). Although they are
normal events, 1937D, 1963P and 1989B 
suffered high extinction in their parent galaxies. We assigned an error of
one magnitude to 1937D and 1963P while
we used the color excess of 1989B ($E(B-V) = 0.37 \pm 0.03$, 
see Branch et al., 1996) in order to correct its magnitude
from total extinction (galactic plus parent-galaxy).
The resulting uncertainty on $m_{B}^{cor.}$ is 0.16 mag. \\
 We also flaged 1971I which has spectral
and light-curve particularities ($\sigma_{m_{B}^{cor.}}$ = 1.00),
 and 1961H which may be overluminous ($\sigma_{m_{B}^{cor.}}$ = 0.22). \\
1963J and 1983U (Tammann \& Sandage, 1995), and 1968E 
(Patat et al., 1997) have very uncertain light curves so that we assigned
 to them $\sigma_{m_{B}^{cor.}}$ = 1.00.  \\
We assigned to 1993af an error $\sigma_{m_{B}^{cor.}}$ = 3.00 because
its spectra showed that it had been caught
several weeks (or months) after maximum luminosity, so that a reliable photometry
was not available (Hamuy et al., 1996b).\\
At last, among our 57 SNe, 7 others have a flag in the catalogue itself
because their real maximum peak brightness may be ``brighter than or equal to''
the plotted magnitudes. These magnitudes refer usually to discovery
which often occured after the peak brightness so 
that $\sigma_{m_{B}^{cor.}}$ = 0.60.\\ 
  
The photometry carried out at Asiago observatory during the seventies is
corrected from errors according to Patat et al (1997); this improvement concerns
1957B, 1960F, 1960R, 1965I, 1970J, 1975N, and Tsvetkov measurements (see also Patat et al., 1997), 1969C,
1971L, 1973N, 1974G and 1974J.\\
We also checked both the total $(B-V)_{T}^{0}$ color of the galaxies (from LEDA, according to de Vaucouleurs et al., 1991) 
to take into account a possible correlation
between color and magnitude peak brightness (Branch et al., 1996), and the 
morphological types to sort out the elliptical galaxies.\\

\begin{figure}
\epsfxsize=8.5cm
\epsfbox[40 400 350 650]{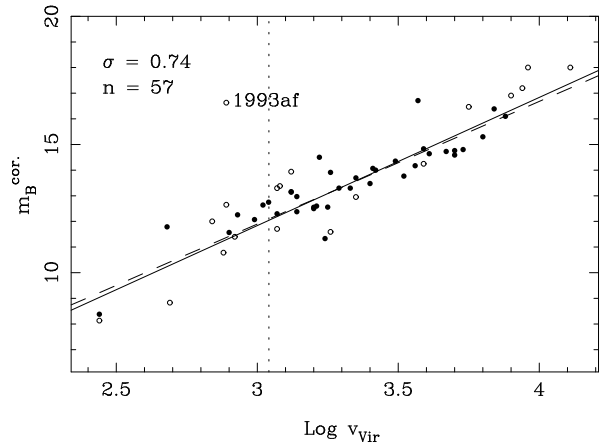}
\caption{Hubble diagram obtained from our sample. The dashed line represents
the best linear fit to a straight line. The continuous line represents 
the best fit, forcing the slope to be the theoretical one. The vertical dotted
line represents the cut in radial velocities to keep those greater than 
$1100~km.s^{-1}$. The filled circles are used for the more confident
SNe Ia, whereas open circles describe SNe  
whose magnitudes are less reliable.
}
\label{hubble-diag}
\end{figure}

We thus obtained a sample of 57 SNe Ia presented in Table \ref{sample} 
that allows us to plot a Hubble diagram  \linebreak (Figure \ref{hubble-diag}).\\ 

\begin{table*}
\begin{tabular}{|llllllll|}
\hline
SNe Ia & Galaxy & R. A. 2000 DEC.& Type & $\log v_{Vir}$ & $m_{B}^{cor.}$ & $\sigma_{m_{B}^{cor.}}$ & $A_{g}$  \\
 & & h mn s deg ' ''& & & & & \\
\hline
 1895 B &NGC5253 &$133955.8-313841$&Sd   &2.436&~8.13!& 0.22 & .20   \\
 1937 C &IC4182  &$130549.3+373621$&Sm   &2.688&~8.83!& 0.15 & .00   \\
 1937 D &NGC1003 &$023916.5+405222$&Sc   &2.891&12.65 & 1.00 & .25   \\   
 1939 A &NGC4636 &$124249.8+024117$&E    &3.037&12.75 & 0.14 & .05   \\
 1957 B &NGC4374 &$122503.7+125315$&E    &2.988&12.07!& 0.14 & .13   \\
 1960 F &NGC4496A&$123139.8+035621$&SBd  &3.241&11.33!& 0.14 & .01   \\
 1960 R &NGC4382 &$122524.6+181127$&SO-a &2.903&11.57!& 0.14 & .03   \\
 1961 H &NGC4564 &$123627.0+112621$&E    &3.069&11.71 & 0.22 & .09   \\
 1963 J &NGC3913 &$115038.9+552112$&Scd  &3.067&13.30 & 1.00 & .00  \\
 1963 P &NGC1084 &$024559.7-073442$&Sc   &3.124&13.94 & 1.00 & .06  \\
 1965 I &NGC4753 &$125222.7-011157$&SO   &3.060&12.37!& 0.14 & .03   \\
 1966 J &NGC3198 &$101954.9+453309$&SBc  &2.916&11.40 & 0.60 & .00   \\
 1967 C &NGC3389 &$104827.8+123201$&Sc   &3.121&13.14 & 0.14 & .06   \\
 1968 E &NGC2713 &$085720.6+025521$&SBab &3.587&14.25 & 1.00 & .15  \\
 1969 C &NGC3811 &$114116.2+474135$&SBc  &3.517&13.77!& 0.14 & .02   \\
 1970 J &NGC7619 &$232014.7+081223$&E    &3.588&14.83!& 0.14 & .17   \\
 1971 I &NGC5055 &$131549.2+420206$&Sbc  &2.841&12.00 & 1.00 & .00  \\
 1971 L &NGC6384 &$173224.5+070338$&SBbc &3.247&12.56!& 0.14 & .44   \\
 1972 E &NGC5253 &$133955.8-313841$&Sd   &2.436&~8.38!& 0.14 & .20   \\
 1973 N &NGC7495 &$230857.3+120254$&Sc   &3.697&14.76!& 0.14 & .15   \\
 1974 G &NGC4414 &$122627.5+311329$&Sc   &2.931&12.26!& 0.14 & .02   \\
 1974 J &NGC7343 &$223837.5+340423$&SBbc &3.884&15.30!& 0.14 & .30   \\
 1975 A &NGC2207 &$061622.0-212221$&SBbc &3.410&14.07 & 0.14 & .53   \\
 1975 N &NGC7723 &$233857.0-125742$&SBb  &3.262&13.91!& 0.14 & .09   \\
 1978 E &MCG+06-49-36&$223459.9+371157$&Sc   &3.704&14.59 & 0.14 & .61   \\
 1979 B &NGC3913 &$115038.9+552112$&Scd  &3.067&12.30 & 0.14 & .00   \\
 1980 N &NGC1316 &$032241.5-371228$&SO   &3.200&12.50 & 0.14 & .00   \\
 1981 B &NGC4536 &$123426.9+021119$&SBbc &3.257&11.59!& 0.23 & .00   \\
 1982 B &NGC2268 &$071415.6+842250$&SBbc &3.396&13.48 & 0.14 & .22   \\
 1982 W &NGC5485 &$140711.5+550008$&SO   &3.221&14.50 & 0.14 & .00   \\
 1983 G &NGC4753 &$125222.7-011157$&SO   &3.060&12.97 & 0.14 & .03   \\
 1983 R &IC1731  &$015012.7+271149$&SBc  &3.557&14.17 & 0.14 & .23   \\
 1983 U &NGC3227 &$102331.4+195148$&SBa  &3.080&13.38 & 1.00 & .02  \\ 
 1983 W &NGC3625 &$112031.7+574655$&SBb  &3.333&13.30 & 0.14 & .00   \\
 1986 A &NGC3367 &$104634.5+134509$&SBc  &3.486&14.35 & 0.14 & .05   \\
 1987 D &MCG+00-32-01&$121940.5+020451$&Sbc  &3.346&13.70 & 0.14 & .00   \\
 1987 N &NGC7606 &$231904.8-082908$&Sb   &3.347&12.95 & 0.60 & .05   \\
 1988 F &MCG+02-37-15a&$142858.6+135142$&SO   &3.730&14.80 & 0.14 & .00   \\
 1989 A &NGC3687 &$112800.6+293041$&SBbc &3.419&14.00 & 0.14 & .00   \\
 1989 B &NGC3627 &$112014.4+125942$&SBb  &2.880&10.78!& 0.16 &       \\
 1989 M &NGC4579 &$123744.1+114911$&SBb  &3.196&12.55 & 0.14 & .15   \\
 1990 N &NGC4639 &$124252.6+131530$&SBbc &3.024&12.64 & 0.14 & .06   \\
 1991 ag&IC4919  &$200009.2-552228$&SBd  &3.611&14.64 & 0.14 & .16   \\
 1991 bb&UGC2892 &$035337.6+190616$&SBbc &3.903&16.91 & 0.60 & .79   \\
 1992 A &NGC1380 &$033626.9-345833$&SO   &3.215&12.60 & 0.14 & .00   \\
 1992 P &IC3690  &$124249.5+102134$&Sbc  &3.884&16.10 & 0.14 & .00   \\
 1992 ap&UGC10430&$163033.2+412936$&SBbc &3.964&18.00 & 0.60 & .00   \\
 1993 I &MCG+2-32-144&$123443.0+090011$&SO  &4.111&18.00 & 0.60 & .00   \\
 1993 ae&UGC1071 &$012944.8-015832$&     &3.754&16.47 & 0.60 & .13   \\
 1993 af&NGC1808 &$050742.7-373051$&SBa  &2.891&16.63 & 3.00 & .07  \\ 
 1993 ah&ESO471-27&$235150.7-275748$&SO  &3.943&17.20 & 0.60 & .00   \\
 1994 D &NGC4526 &$123402.9+074201$&SO   &2.682&11.79 & 0.14 & .01   \\
 1994 M &NGC4493 &$123108.5+003648$&E    &3.841&16.39 & 0.14 & .01   \\
 1994 S &NGC4495 &$123123.1+290813$&Sab  &3.670&14.73 & 0.14 & .07   \\
 1994 ae&NGC3370 &$104703.6+171626$&Sc   &3.122&13.16 & 0.14 & .04   \\
 1995 D &NGC2962 &$094054.0+051000$&SO-a &3.286&13.30 & 0.14 & .10   \\
 1995 E &NGC2441 &$075154.6+730058$&SBb  &3.569&16.71 & 0.14 & .09   \\
\hline
\hline
\end{tabular}
\caption{Sample of 57 Supernovae of type Ia. {\it Column 1} : SNe name. 
{\it Column 2} : Parent galaxy name. {\it Column 3} : Equatorial
coordinates for equinoxe 2000 of parents galaxies. {\it Column 4} : Parent
galaxy morphological type.
{\it Column 5} : $Log_{10}$ of the radial velocity corrected from the
Virgo infall. {\it Column 6 and 7} : Apparent B magnitude corrected for
extinction and the associated uncertainty. The flag (!) 
means that the value has been modified from
the Asiago catalogue value. 
{\it Column 8} : Galactic extinction (B-band) according to 
Burstein \& Heiles (1984).
We don't list $A_{g}$ for NGC 3627, because we calculated the galactic 
extinction from the color excess. 
}
\label{sample}
\end{table*}

\subsection{Analysis}
If we consider a linear expansion model, the Hubble diagram will be best
fitted by the law :
\begin{equation}
 m_{B}^{cor.} = 5 \log v_{Vir} + ZP_{B}
\label{hubble}
\end{equation}
The uncertainties on $\log v_{Vir}$ are small enough (less than a few hundreths)
 to satisfy the following condition :
\begin{eqnarray}
5~\sigma_{\log v_{Vir}} = \frac{5}{\ln10} \frac{\sigma_{v_{Vir}}}{v_{Vir}} \ll \sigma_{m_{B}^{cor.}} \nonumber
\end{eqnarray}
 which allows us to
use a direct regression. \\ 
If we then use weighted measurements of our sample of 57 SNe Ia (the weight
being defined as $1/\sigma_{m_{B}^{cor.}}^2$), 
we obtain  from a maximum likelihood \linebreak method :
\begin{eqnarray}
m_{B}^{cor.}& = &(4.779 \pm 0.064) \log v_{Vir} + (-2.44 \pm 0.21) \\
\sigma & = & 0.74  \nonumber
\label{libre57}
\end{eqnarray}
A Student's t-test on the slope leads to :                            
\begin{equation}                          
t = \frac{|4.779 - 5|}{0.064}= 3.45    
\end{equation}
which is smaller than $t_{0.01}$ (the probability of error is much lower
than 0.01). We conclude that the slope is not significantly different from 
the theoretical one at this probability level. \\

\begin{figure}
\epsfxsize=8.5cm
\epsfbox[40 400 350 650]{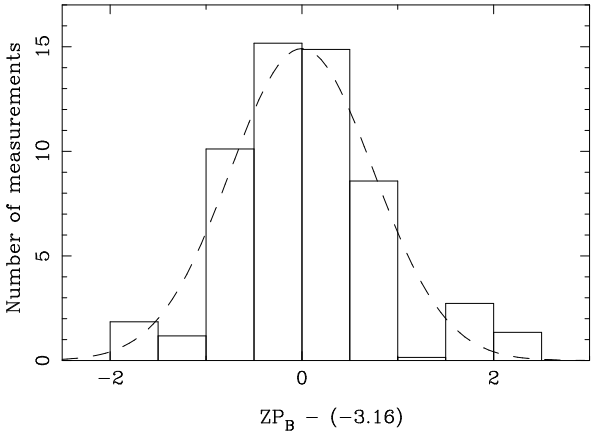}
\caption{Histogram of the weighted difference $(ZP_{B} - (-3.16))$. The 
dashed Gaussian curve was calculated from the mean and standard deviation
estimated from these measurements. The $ZP_{B}$ obtained from SN 1993af
was excluded from this figure.
}
\label{zp}
\end{figure}
Then forcing the slope to be exactly 5, we obtain :
\begin{equation}
ZP_{B} = -3.16 \pm 0.10 
\label{force57}   
\end{equation}
In order to test if the distribution is normal, we plot the 
weighted histogram of the difference
$(ZP_{B} - (-3.16))$ (see Figure \ref{zp}).
We perform a $\chi^{2}$ test for this distribution and we obtain 
$\chi^{2} = 1.84$. Referring to a $\chi^{2}$ table, we observe that, for 8
degrees of freedom, the probability of obtaining in repeated experiments 
a greater value of $\chi^{2}$ is $\sim$ 8\%, so that we can consider this
distribution to be Gaussian ($<ZP_{B} - (-3.16)> = 0, \sigma = 0.74)$.\\      

If we sort out the SNe in our sample according to the 
nature of their parent galaxy (elliptical or not),
it comes down to excluding five events
that occured in elliptical galaxies (only 1939A, 1957B, 1961H, 1970J and 1994M 
because 91\% of our sample is made of SNe Ia that occured in spiral galaxies)
and we arrive at the same value of $ZP_{B}$ within about three hundreths
($ZP_{B} = -3.20 \pm 0.11$).   \\
At last, we can sort out the sample according to the total color of the parent
 galaxy $(B-V)_{T}^{0}$. If we keep only those with $(B-V)_{T}^{0} \leq 0.75$,
we obtain \linebreak $ZP_{B} = -3.37 \pm 0.15 (n=25)$. The 13 remaining SNe Ia, with
$(B-V)_{T}^{0} \geq 0.75$, leading us to $ZP_{B} = -2.83 \pm 0.21 
$ (note that 19 galaxies have no measurement available).
These two results tend to show that SNe Ia in redder galaxies have lower
luminosities in agreement with Branch et al. results (1996), but 
following these authors, we won't take this effect further into account when
we will compute the Hubble constant. \\
If we only keep the velocities above $1100~km.s^{-1}$ 
to \linebreak minimize the particular motions among the 57 previous SNe,
we obtain a sample of 44 SNe Ia (note that none of the velocity are high
enough to justify a departure from linearity in Hubble's law) and :
$ZP_{B} = -3.19 \pm 0.11$. This result does not differ significantly from 
the main value.\\

\subsection{The value of $H_{0}$}
Let us now recall the relation :
\begin{equation}
m - M = 5 \log d_{Mpc} + 25
\label{mu}
\end{equation}
From  the equations \ref{hubble}, \ref{mu} and the Hubble approximation
 $ H_{0} \approx v_{Vir} / d_{Mpc} $
 we obtain :
\begin{equation}
\log H_{0} = 0.2 \langle M_{B}(MAX) \rangle + 5 - 0.2~ZP_{B}
\label{  }
\end{equation}
If we then use the best value $ ZP_{B} = -3.16 \pm 0.10$ 
(equation \ref{force57}) and 
$ \langle M_{B}(MAX) \rangle = -19.65 \pm 0.09$ (equation \ref{Mmean}), it 
leads us to (assuming $\sigma_{H_{0}} = H_{0}~\ln 10~\sigma_{\log H_{0}}$) : 
\begin{equation}
 H_{0} = 50 \pm 3~(internal)~km.s^{-1}.Mpc^{-1}
\label{resultat}
\end{equation}
If we consider the external errors, that is especially $ \langle M_{B}(MAX) \rangle = -19.66 \pm 0.18$ (equation \ref{Mmean2}), we obtain~: 
\begin{equation}
 H_{0} = 50 \pm 5~(external)~km.s^{-1}.Mpc^{-1}
\label{resultat}
\end{equation}

We have to keep in mind that our present goal is mainly to recalibrate
the maximum peak brightness thanks to the HIPPARCOS data, and not to
build a new Hubble diagram. Thus we can also use other diagrams such
as presented by Saha et al. (1997), and recalibrate it
thanks to our new peak calibration. Using their zero-point 
$ZP_{B} = -3.265 \pm 0.045$, we would obtain
$H_{0} = 53 \pm 2~(internal)~km.s^{-1}.Mpc^{-1}$.\\

The same application to  the Tammann \& Sandage's diagram (1995) with their 
$ZP_{B} = -3.186 \pm 0.054$ leads to 
$H_{0} = 51 \pm 2~(internal)~km.s^{-1}.Mpc^{-1}$.\\

At last, using the value $ZP_{B} = -3.177 \pm 0.029  $ from Hamuy et al.
(1996a), we compute 
$H_{0} = 51 \pm 2~(internal)~km.s^{-1}.Mpc^{-1}$ (and taking into account their
absolute magnitude-decline rate relation, we derive from : 
$H_{0} = 54 \pm 2~km.s^{-1}.Mpc^{-1}$). \\
Note that the external errors in the three previous applications are about $4~km.s^{-1}.Mpc^{-1}$.\\

Table \ref{dependance} puts together the various results we computed. 
It obviously appears that these results depend very slightly on the choices
we made.\\ 
 
\begin{table}
\begin{tabular}{|l|l|l|l|l|l|l|}
\hline
 &-3.16&-3.265&-3.186&-3.177&-3.318 \\
 & (1) & (2) & (3) & (4) & (4) \\
\hline
-19.65 (a) &50   &53   &51   &51   &54 \\
-19.62 (b) &51   &54   &52   &51   &55 \\         
\hline
\hline
\end{tabular}
\caption{Dependance of $H_{0}$ on our choices (couples $(\langle M_{B}(MAX)
 \rangle,ZP_{B}))$. (1) This paper. (2) Saha et al., 1997. (3) Tammann 
and Sandage, 1995. (4) Hamuy et al., 1996a. (a) Main weighted value from the 
four calibrators. (b) Value obtained from 1937C and 1972E only.}
\label{dependance}
\end{table}

\section{Conclusion}
The main observation is that the calibration from \linebreak HIPPARCOS data favors
 the so called ``long distance scale'', and is in great agreement with
Saha and coworkers's who obtained $H_{0} = 58_{-8}^{+7}~km.s^{-1}.Mpc^{-1}$
in their last paper (1997). \\
We could also consider the effect of the decline rate ($\Delta m_{15}$) on
the absolute magnitude and, therefore, on $H_{0}$ (Phillips, 1993). However this effect
seems to be not yet well determined, and would however increase $H_{0}$
by less than 10 \% (Saha et al., 1997), even if we 
take into account others effect such as
SNe Ia color or Hubble type of the parent galaxy.\\
We also note that the present work confirms another result based on the same 
HIPPARCOS calibration (Paturel et al., 1997c), where we obtained 
$H_{0} = 53_{-8}^{+7}~km.s^{-1}.Mpc^{-1}$ through a completely independent
way.\\ 
We must add that, although a part of the statistical bias on calibrating
distance moduli was corrected, the present moduli could still be affected by a residual
bias (Paturel et al., 1997a). 
In that case, the correction needed would induce a lower value for $H_{0}$.\\
 
\acknowledgements{I would like to thank Dr. G. Paturel and O. Witasse for their help. I also want to thank the referee for its useful comments.\\
We have made use of the Lyon-Meudon Extragalactic Database                   
(LEDA) supplied by the LEDA team at the CRAL-Observatoire de                   
Lyon (France). 
}


\begin{thebibliography}{}
\bibitem{} Aksenov, E. P., 1981 IAU Circ., 3580
\bibitem{} Ardeberg, A., de Groot, M., 1973, A \& A, 28, 295
\bibitem{} Arp, H. C., 1961, ApJ, 133, 871                 
\bibitem{} Baade, W., Zwicky, F., 1938, ApJ, 88, 411   
\bibitem{} Barbon, R., Cappellaro, E., Turatto, M., 1989, A\&ASS, 81, 421
\bibitem{} Branch, D., Lacy, C. H., McCall, M. L., Sutherland, P. G., Uomoto, A., Wheeler, J. C., Wills, B. J., 1983, ApJ, 270, 123 
\bibitem{} Branch, D., Miller, D. L., 1993, ApJ, 405, L5 
\bibitem{} Branch, D., Romanishin, W., Baron, E., 1996, ApJ, 465, 73
\bibitem{} Burstein, D., Heiles, C., 1984, ApJ S., 54, 33    
\bibitem{} Buta, R. J., Turner, A., 1983, PASP, 95, 72       
\bibitem{} Fleming, W. P., Pickering, E. C., 1896, ApJ, 3, 162
\bibitem{} Hamuy, M., Phillips, M. M., Suntzeff, N. B., Schommer, M. M., Maza, R. A., Aviles, R., 1996a, AJ, 112, 2398
\bibitem{} Hamuy, M., Phillips, M. M., Suntzeff, N. B., Schommer, M. M., Maza, R., Antezan, A. R., Wischnjewsky, M.,
Valladares, G., Muena, C., Gonzales, L. E., Aviles, R., Wells, L. A., Smith, R. C., Navarrete, M.,
Covarrubias, R., Williger, G. M., Walker, A. R., Layden A. C., Elias, J. H., Baldwin, J. A., Hernandez, M.,
Tirado, H., Ugarte, P., Elsion, R., Saavedra, N., Barrientos, F., Costa, E., Lira, P., Rutz, M. T., Anguita, C.,
Gomez, X., Ortiz, P., Della Valle, M., Danziger, J., Storm, J., Kim Y-C., Bailyn, C., Rubenstein E . P.,
Tucker, D., Cersosimo, S., Mendez, R. A., Siciliano, L., Sherry, W., Chaboyer, B., Koopmann, R. A.,
Geisler, D., Sarajedini, A., Dey, A., Tyson, N., Rich, R. M., Gal, R., Lamontagne, R., Caldwell, N.,
Guhathakurta, P., Phillips, A. C., Szkody, P., Prosser, C., Ho, L. C., McMahan, R., Baggley, G., Cheng, K. P.,
Havlen, R., Wakamatsu, K., Jane S, K., Malkan, M., Baganoff, F., Seitzer, P., Shara, M., Sturch, C.,
Hesser, J., Hartigan, P., Hugues, J., Welch, D., Williams, T. B., Ferguson, H., Francis, P. J., French, L.,
Bolte, M., Roth, J., Odewahn, S., Howell, S., Krzeminski, W., 1996b, AJ, 112, 2408
\bibitem{} Hubble, E., Lundmark, K., 1922, PASP, 34, 292       
\bibitem{} Kowal, C. T., 1968, AJ, 73, 1021    
\bibitem{} Kowal, C. T., 1972, IAU Circ., 2405          
\bibitem{} Leibundgut, B., Tammann, G. A., 1990, A\&A, 230, 81 
\bibitem{} Leibundgut, B., Tammann, G. A., Cadonau, R., Cerrito, D., 1991, A\&AS, 89, 357
\bibitem{} Patat, F., Barbon, R., Cappellaro, E., Turatto, M., 1997, A\&A, 317, 423 
\bibitem{} Paturel, G., Lanoix, P., Garnier, R., Rousseau, J., 
 1997a, ESA Symposium : ``HIPPARCOS Venice 1997'', eds Bernacca P. L., Perryman, M. A. C., ESA SP-402
\bibitem{} Paturel, G., Bottinelli, L., Di Nella, H., Durand, N., Garnier, R.,
Gouguenheim, L., Lanoix, P., Marthinet, M. C., Petit, C., Rousseau, J., 
Theureau, G., Vauglin, I., 1997b, A\&A, in press
\bibitem{} Paturel, G., Bottinelli, L., Gouguenheim, L., Lanoix, P.,   
Renaud, N., Teerikorpi, P., Theureau, G., Witasse, O., 1997c, A\&A, submitted   
\bibitem{} Phillips, M. M., 1993, ApJ Letters, 413, L105                       
\bibitem{} Saha, A., Sandage, A., Labhardt, L., Schwengler, H., Tammann, G. A.,
Panagia, N., Machetto F. D., 1994, ApJ, 425, 14
\bibitem{} Saha, A., Sandage, A., Labhardt, L., Schwengler, H., Tammann, G. A.,
Panagia, N., Machetto F. D., 1995, ApJ, 438, 8
\bibitem{} Saha, A., Sandage, A., Labhardt, H., Tammann, G. A.,
Machetto F. D., Panagia, N., 1996, ApJ, 466, 55
\bibitem{} Saha, A., Sandage, A., Labhardt, H., Tammann, G. A.,
Machetto F. D., Panagia, N., 1997, preprint              
\bibitem{} Sandage, A., Tammann, G. A., 1990, ApJ, 365, 1
\bibitem{} Sandage, A., Tammann, G. A., 1993, ApJ, 415, 1
\bibitem{} Savage, B. D., Mathis, J. S., 1979, Ann. Rev. Astr. Astrophys.
17, 73
\bibitem{} Schaefer, B. E., Bradley, E., 1995, ApJ Letters, 447, L13  
\bibitem{} Schaefer, B. E., 1995, ApJ Letters, 449, L9                
\bibitem{} Tammann, G. A., Sandage, A., 1995, ApJ, 452, 16                
\bibitem{} Vaucouleurs, G. de, Vaucouleurs, A. de, Corwin, H. G., Buta, R. J., Paturel, G., Fouqu\'e, P., 1991, Third Reference Catalogue of Bright Galaxies, Springer-Verlag (RC3) 
\bibitem{} Walker, A., D., 1923, Harvard Ann., 84, 189
\end{thebibliography}
\end{document}